\documentclass[12pt]{article}
\usepackage{amsmath}
\usepackage{amssymb}
\usepackage{indentfirst}
\usepackage{amssymb}
\usepackage{amsfonts}
\usepackage{amscd}
\usepackage{amsbsy}
\usepackage{amsthm}
\usepackage{latexsym}
\usepackage{graphicx,color} %,epsf}
\usepackage[dvipsnames]{xcolor}
\usepackage[colorlinks]{hyperref}
\hypersetup{linkcolor=blue,citecolor=blue,urlcolor=blue}

\def\nn{\nonumber}       %%%    nonumber
\def\beq{\begin{eqnarray}}
\def\eeq{\end{eqnarray}}

\def\diag{\,\mbox{diag}\,}

\def\al{\alpha}
\def\be{\beta}

\def\ga{\gamma}

\def\ep{\epsilon}

\def\la{\lambda}

\def\pa{\partial}

\def\si{\sigma}

\def\ph{\varphi}

\def\De{\Delta}

\begin{document}

\begin{center}
%%%%%%%%%%%%%%%%%%%%%%%%%%%%%%%%%%

{\Large % \bf
Inflation with sterile scalar coupled to massive fermions and to gravity}
\vskip 6mm

\textbf{J\'{e}ssica~S.~Martins}$^{(a)}$\footnote{Email: \ jessica.martins@unesp.br},
\ \
\textbf{Oliver~F.~Piattella}$^{(b,c)}$\footnote{Email: \ of.piattella@uninsubria.it},
\ \
\textbf{Ilya~L.~Shapiro}$^{(d,e,f)}$\footnote{Email: \ ilyashapiro2016@gmail.com},
\\ \textbf{Alexei~A.~Starobinsky}$^{(g,h)}$\footnote{Email: \ alstar@landau.ac.ru}
\vskip 6mm

$^{(a)}$ \ Instituto de F\'isica Te\'orica,
Universidade Estadual Paulista - Campus de S\~ao Paulo, 01140-070 S\~ao Paulo, Brazil
\vskip 2mm

$^{(b)}$ N\'ucleo Cosmo-UFES \& Departamento de F\'isica,
Universidade Federal do Esp\'irito Santo, Avenida Fernando Ferrari
514, 29075-910 Vit\'oria, Esp\'irito Santo, Brazil
\vskip 2mm

$^{(c)}$ Dipartimento di Scienza ed Alta Tecnologia, Universit\`a dell'Insubria, Via Valleggio 11, I-22100 Como, Italy
\vskip 2mm

$^{(d)}$ Departamento de F\'{\i}sica, ICE, Universidade Federal de Juiz de Fora
\\
Campus Universit\'{a}rio - Juiz de Fora, 36036-330, Minas Gerais, Brazil
\vskip 2mm

$^{(e)}$ Tomsk State Pedagogical University, 634061, Tomsk, Russia

$^{(f)}$ Tomsk State University, 634050, Tomsk, Russia

$^{(g)}$ L. D. Landau Institute for Theoretical Physics RAS,
Moscow, 119334, Russia
\vskip 2mm

$^{(h)}$ National Research University Higher School of Economics,
Moscow, 101000, Russia
\vskip 2mm

\end{center}
%%%%%%%%%%%%%%%%%%%%%%%%%%%%%%%%%%

%\begin{history}
%\received{Day Month Year}
%\revised{Day Month Year}
%\end{history}

\begin{abstract}
\noindent
In the recent paper \cite{AJV} it was shown that the consistency of
the quantum theory of a sterile scalar coupled to massive fermions
requires the inclusion of odd-power terms in the potential of scalar
self-interaction. One of the most important examples of a sterile
scalar is the inflaton, that is typically a real scalar field which
does not belong to representations of particle physics gauge groups,
such as $SU(2)$. Here we explore the effects of the odd-power terms
in the inflation potential on main observables, such as the scalar
spectral index $n_s$ and the tensor-to-scalar ratio $r$, in the case of
inflaton strong non-minimal coupling to gravity.
\vskip 3mm


%%%%%%%%%%%%%%%%%%%%%%%%%%%%%%%%
\end{abstract}

%\ccode{PACS numbers:}
%\tableofcontents

%%%%%%%%%%%%%%%%%%%%%%%%%%%%%%%%
%%%%%%%%%%%%%%%%%%%%%%%%%%%%%%%%
%%%%%%%%%%%%%%%%%%%%%%%%%%%%%%%%
\section{Introduction}
\label{sec1}

Inflaton-based models play an important role in the understanding of inflation. There is an extensive variety of the inflaton models \cite{Martin:2013nzq}, however typically the inflaton is a specially designed real scalar field with the potential providing a special dynamics of the vacuum, such that the Universe can inflate in a proportion required by existing observational data. Since the inflaton is a scalar which is not a representation of a gauge group of the Standard Model, it can be called a sterile scalar. On the other hand, this scalar has to be coupled to ordinary matter in order for the reheating phase to take place at the end of the inflationary period \cite{Kofman:1994rk}.

Recently, it was shown that a sterile scalar coupled to massive
fermions has to satisfy certain consistency conditions, related to
quantum corrections  \cite{AJV}. In particular, the renormalizability
of such theory can be achieved only if the inflaton potential is
supplemented by three terms which have odd powers of the sterile
scalar field. A relevant detail is that these odd-power terms are not
necessary in other models e.g. in the Higgs inflation \cite{Shaposh}.
In this model the loop corrections are also important, as they define
the value of the non-minimal parameter $\xi$ and even impose the
constraints on the Higgs mass (see e.g.
\cite{Barvinsky:2008ia, Barvinsky:2009fy} and
\cite{Bezrukov:2008ej, Bezrukov:2010jz, Bezrukov:2009db}), but
there is no need to include odd terms, since the Higgs field is not a
sterile field.

The situation described above opens the following interesting possibility. Up to some extent, the inflaton-based models can be mapped to the $f(R)$-type modified theory of gravity. In general, this requires a conformal transformation (or even two of them in case of a non-minimal coupling of inflaton to gravity), but in the case of strong non-minimal coupling, this can be achieved for inflationary trajectories in the phase space even without using it. This was done for example in~\cite{Miranda:2017juz} for the case of the $\alpha$-Attractors class of inflationary models \cite{Kallosh:2013hoa}, and in~\cite{He:2018gyf} for the mixed Higgs-$R^2$ inflationary model. This feature certainly remains valid for the inflaton models with odd potential. But if the inflaton is a real field, after such a mapping we shall meet very specific additions to the function $f(R)$ that may produce observables which are different from the ones of other, most frequently used, functions. As far as odd terms in the potential are typical only for the inflaton-based models, one can use the observable consequences of these terms to learn whether inflation is caused by the inflaton, or by some form of modified gravity theory.

In the present work we shall explore this possibility. To do so,
we use the following strategy. We map the inflaton potential with odd terms to a $f(R)$ gravity model without a conformal transformation, i.e. staying in the same Jordan frame, that can be done approximately in the case of strong non-minimal inflaton coupling to gravity and sufficiently smooth behaviour of the inflaton like that during slow-roll inflation. Thus, particle masses, the Hubble parameter and space-time curvature keep their original physical values during this mapping. The corresponding function $f(R)$ has additional terms due to the odd powers in the inflaton potential, and further work will concern these additional terms. Since the odd terms are assumed to be numerically small, the resulting $f(R)$ will be a sum of the usual $R^2$-term, typical of the Starobinsky model of inflation \cite{star, star83}, plus extra terms which produce the effect of our interest.

The paper is organized as follows. In Sec.~\ref{sec2.0} we use the perturbation theory and conformal transformation to perform the mapping of the original potential with odd terms \cite{AJV} to the Einstein frame, that is the most standard way for the analysis of inflationary parameters. An alternative mapping without the use of a conformal transformation is considered in Sec.~\ref{sec2}. It leads to the $f(R)$ model having the same inflationary stage with the same predictions for primordial perturbation spectra as the original model. In Sec.~\ref{s3} we derive and analyse the inflationary slow-roll parameters. Finally, in Sec.~\ref{sec5} we draw our conclusions.

%%%%%%%%%%%%%%%%%%%%%%%%%%%%%%%%%%%
\section{Scalar field with odd terms and transformation
to the Einstein frame}
\label{sec2.0}

Assuming that the inflaton (or other kinds of sterile scalar) $\ph$
couples to fermions $\psi_k$ by means of a Yukawa-type interaction,
i.e., $h_k {\bar \psi}_k \ph \psi_k$, the picture is qualitatively
different from the one for the fermion-Higgs interactions. The
Higgs scalar belongs to the fundamental representation of the
$SU(2)$ gauge group and therefore has the corresponding group
index, $\Phi = \Phi_i$, with $i = 1,2$. As a result, the divergences
of the odd powers of $\Phi$ are forbidden. For a sterile scalar this
is not the case and one can expect that the corresponding divergences
show up, according to power counting. The explicit calculations
using the heat-kernel method \cite{AJV} have shown that the
corresponding counterterms emerge already at the one-loop level.
According to standard arguments, this means that the odd terms
should be included already at the classical level, in order for the
theory to be renormalizable. If we do not follow this standard
procedure, the odd terms will emerge anyway, proportional to the
leading logarithms in momenta or scalar field, and will be more
difficult to control.

Taking into account the non-minimal interaction between the sterile
scalar field and the scalar curvature, the potential of the sterile
scalar reads
\beq
V(\varphi)
\,=\,
\frac{1}{2}m^2\varphi^2 + \frac{\lambda}{4!}\varphi^4
- \frac12 \xi\varphi^2 R
+ \frac{g}{3!}\varphi^3 - \tau\varphi + \tilde g R\varphi,
\label{Vcla}
\eeq
where $\la$ is the usual dimensionless scalar self-coupling
parameter and $m$ is the scalar mass.

We are using the notations $\eta_{\mu\nu} = \diag(-1,1,1,1)$. As a
consequence, $R>0$ during inflation, and so $\xi>0$ is the necessary
condition for the non-minimal coupling. In fact, we assume strong non-minimal
coupling, $\xi\gg 1$, similar to what is required for the Higgs inflationary model.
On the top of this, we need
$\la>0$ for the stability of the vacuum state. Furthermore, $g$, $\tau$
and $\la$ are the non-minimal parameters corresponding to the odd
powers of the scalar. Different from $\la$ and $\xi$ these parameters
are dimensional, $[g]= [mass]$, $[\tilde g]=[mass]$ and $[\tau]=[mass^3]$.

The analysis of the renormalization group equations for $g$, $\tau$
and $\tilde g$ shows that the minimal possible magnitudes of these
parameters are defined by the masses of heaviest fermions, i.e. the
top quark, in the Standard Model. Compared to the value of the
Hubble parameter, even at the end of inflation, these values are
small. However, even relatively small parameters can produce
measurable effects if the corresponding terms are qualitatively
different. Thus, in what follows we shall try to explore such traces
for the odd terms in the potential (\ref{Vcla}).

As a first step, let us transform the action of gravity and non-minimal
scalar with potential (\ref{Vcla}),
\beq
S\,=\,\int d^4x \sqrt{-g}\bigg\{\frac{M^2_P}{2}R
\,+\,
\frac12\,g^{\mu\nu}\pa_\mu \ph \pa_\nu \ph - V(\varphi) \bigg\},
\label{Scla}
\eeq
into the Einstein frame. We assume that the terms
$V_0(\ph) = \frac{\la}{4!}\ph^4 - \frac12 \xi R\ph^2$
are dominating and treat the rest of the potential,
$V_1(\ph) = \frac{1}{2}m^2\ph^2
+ \frac{g}{3!}\ph^3 - \tau\ph + {\tilde g} R\ph$,
as a small perturbation, taking only first order effects into account.
Also, during the inflationary epoch, the kinetic term can be
neglected, so we shall not take it into account, even as a
perturbation. We shall denote the solution of the corresponding
equation of motion with $V_0(\ph) $ as $\ph_0$, and the solution of
the full equation, with $V(\ph)=V_0(\ph)+V_1(\ph)$, as $\ph_0+\ph_1$.

First we consider the theory with the basic potential $V_0(\ph)$.
It proves useful to perform the following change of variables
in the zero-order action:
\beq
M_P^2 B \,=\,\xi \ph_0^2,
\label{change0}
\eeq
where $B$ is a new scalar field. Then the reduced (without the
kinetic term) form of the action (\ref{Scla}) is
\beq
S_0\,=\,\int d^4x \sqrt{-g}\bigg\{ \frac{M_P^2}{2}(B+1) R
- \frac{M_P^4}{8\al}B^2
\bigg\},
\label{SclaB0}
\eeq
where $\al$ is the first of the useful new parameters
\beq
\al = \frac{3\xi^2}{\la},
\qquad
\be =  \tilde g + \frac{g\xi}{\la},
\qquad
\ga = \sqrt{\frac{2\al}{\xi}}.
\label{albega}
\eeq
Note that the parameter $\be$ is invariant under an arbitrary shift in $\ph$: $\ph \to \ph + \delta\ph$.

Making the conformal transformation of the metric (not of the
scalar field)
\beq
g_{\mu\nu}\,=\,{\tilde g}_{\mu\nu}e^{2\rho},
\qquad
e^{2\rho} = 1+B\,,
\label{rho}
\eeq
after some algebra we arrive at the action
\beq
S_0
\,=\,
\int d^4x \sqrt{-{\tilde g}}\,\bigg\{
\frac{M_P^2}{2}{\tilde R}
- \frac12\,{\tilde g}^{\mu\nu}\pa_\mu \chi \pa_\nu \chi
- U_0(\chi) \bigg\},
\label{SclaB01}
\eeq
and the minimal potential term is
\beq
U_0(\chi) \,=\,\frac{M^4_P}{8\al}\,\si^2 B^2,
\label{Uchi0}
\eeq
written in terms of the variables
\beq
\sigma \,=\, \sigma(\chi)
\,=\, e^{-\sqrt{\frac{2}{3}}\frac{\chi}{M_P}}\,,
\qquad
B \,=\, B(\chi) \,=\, e^{\sqrt{\frac{2}{3}} \frac{\chi}{M_P}} - 1.
\label{confa}
\eeq
This potential is the Einstein-frame mapping of the $R+R^2$ action
of the Starobinsky inflationary model.

As the next step, consider the first order in perturbations. Starting
from the modified version of the change of variables (\ref{change0}),
we get
\beq
M_P^2 B \,=\,\xi \ph^2 - 2{\tilde g}\ph.
\label{change1}
\eeq
Replacing $\ph = \ph_0 + \ph_1$ into this equation, in the first
order in ${\tilde g}$, we get
\beq
\ph^2 \,=\, \frac{M_P^2 B}{\xi}
\,+\,\frac{2{\tilde g} M_P B^{1/2}}{\xi^{3/2}}.
\label{change2}
\eeq
The action in terms of the field $B$ has the form
\beq
S_1\,=\,
\,=\,\int d^4x \sqrt{-g}\bigg\{ \frac{M_P^2}{2}R(B+1)
- \frac{\la M_P^4 B^2}{24\,\xi^2} - V_1 (B) \bigg\},
\label{SclaB1}
\eeq
where
\beq
V_1 (B)
\,=\,
\frac{m^2 M_P^2 B}{2\xi}
\,+\,
\frac{g M_P^3 B^{3/2}}{\xi^{3/2}}
\,+\,
\frac{{\tilde g}\la M_P^3 B^{3/2}}{6\xi^{5/2}}
\,-\,
\frac{\tau M_P B^{1/2}}{\xi^{1/2}}.
\label{Pot1}
\eeq

Before making the conformal transformation, let us write the
action (\ref{SclaB1}) in terms of a more useful notation. From
now on, we shall express the parameters in units of the (reduced)
Planck mass $M_P=(8\pi G)^{-1/2}$, that is
\beq
m' = \frac{m}{M_P},
\qquad
\tau' = \frac{\tau}{M^3_P},
\qquad
\tilde g' = \frac{\tilde g}{M_P},
\qquad
g' = \frac{g}{M_P}.
\eeq
In these units, according to what we have discussed above,
$m',\,\vert \tau'\vert ,\,\vert  \tilde g'\vert ,\, \vert g'\vert  \ll 1$.

It is important that the conformal transformation is not affected
by the small perturbation terms, and should have the same form
(\ref{confa}) is in the unperturbed version of the theory. The
reason is that the curvature $R$ enters Eq.~ (\ref{SclaB1}) in
exactly the same form as in the action  (\ref{SclaB0}), so this
is the unique form of conformal transformation providing the
canonical kinetic term for the scalar $\chi$.

Dropping the primes,  the new action has the form (\ref{SclaB01})
with the full potential of the form
\beq
U(\chi)
&=&
\frac{M^4_P}{8\al}\sigma^2 B^2
\,+\,  \frac{M^2_P m^2}{2\xi}\sigma^2 B
\,+\,  \frac{\sqrt{2} \ga \be M^4_P}{4\al^{3/2}}
\sigma^2 B^{3/2}
\,-\, \frac{\tau \ga M^4_P}{\sqrt{2\al}}
\si^2 B^{1/2},
\qquad
\label{potentialUchi}
\eeq
As it should be expected, Eq.~\eqref{potential} includes the
potential corresponding to the $R+R^2$ model, plus a perturbation.

It is remarkable that the small parameters $g$, $\la$, $\tilde g$,
and $\tau$ and the parameters $\la$ and $\xi$, enter the expression
(\ref{potentialUchi}) only in the combinations $\al$, $\be$, $\ga$
and $\tau$, the first three defined in (\ref{albega}). The scalar
field $\chi$ combine into the quantities defined in (\ref{confa}).

The derivative of the above potential \eqref{potentialUchi} is the
following:
\begin{eqnarray}
&&
U'(\chi) = \frac{M^3_P}{2\sqrt{6}\al}\si^2 B
+ \frac{M_P m^2}{\sqrt{6}\xi}\sigma(2\sigma - 1)
\nn
\\
&&
\qquad
\qquad
+ \,
\frac{\gamma\beta M^3_P}{4\sqrt{3}\al^{3/2}}\si B^{1/2}
\left(3 - 4\si B \right)
- \frac{\tau\ga M^3_P}{2\sqrt{3\al}}\frac{\sigma}{B^{1/2}}
\left(1 - 4\si B \right)\;.
\end{eqnarray}
In this expression, the first term is the usual one in the Starobinsky
model. Now, for $\chi \to \infty$, we have that $\sigma \to 0$.
Noting that $\sigma B = 1 - \sigma$ and keeping the leading orders
in $\sigma$, in this limit we have then:
\begin{eqnarray}
U'(\chi) \sim \frac{M^3_P}{2\sqrt{6}\al}\si
+ \frac{M_P m^2}{\sqrt{6}\xi}\sigma(-1)
+ \frac{\gamma\beta M^3_P}{4\sqrt{3}\al^{3/2}}\si^{1/2}
\left(-1\right) - \frac{\tau\ga M^3_P}{2\sqrt{3\al}}\sigma^{3/2}
\left(-3\right)\;.
\end{eqnarray}
 It is easy to see that $U' \to 0$ in the limit $\chi \to 0$. Indeed, there
 are various contributions, the second term being of the same order
 $\sigma$ as for the Starobinsky model, plus the third one, which is
 of order $\sigma^{1/2}$ and, therefore, dominant (of course, not
 taking into account the smallness of the parameters in the
 coefficients) and the last one which is subdominant. Note that
 $U'$ must be positive, in order to allow for the slow-roll phase
 towards small values of $\chi$, so we must have, roughly speaking,
 that:
\begin{equation}
	m^2/\xi \ll M_P^2/\alpha\;,
\qquad
\gamma\beta \ll \alpha\;, \qquad \tau\gamma \ll \sqrt{\alpha}\;.
\end{equation}

%%%%%%%%%%%%%%%%%%%%%%%%%%%%%%%%%%%
%%%%%%%%%%%%%%%%%%%%%%%%%%%%%%%%%%%
\section{Induced action of gravity with odd terms}
\label{sec2}

Another way of obtaining the results of the previous
section, which is even simpler in fact, is to use the possibility of the approximate representation of the theory (\ref{Scla}) with $\xi\gg 1$ as $f(R)$ gravity in the same Jordan frame (i.e. without a conformal transformation) up to small terms $\propto \xi^{-1}$. This possibility follows already from the fact that the effective Brans-Dicke parameter
$\omega_{BH}$ %% =F(\phi)/(dF/d\phi)^2$
%%  $\omega_{BH}=F(\phi)/(dF/d\phi)^2$
is very small ($\approx \frac{1}{4\xi}$) for this theory while it is exactly zero for $f(R)$ gravity. The alternative derivation presented below demonstrates the possibility to avoid conformal transformation and to work directly in the Jordan frame all the time. It is useful in the case of large fermion masses since neither particle rest masses, nor the physical values of the Hubble function $H(t)$ are invariant under the conformal transformation.

Our strategy will be as follows. We perform mapping of the scalar theory with the potential (\ref{Vcla}) strongly coupled to the Ricci scalar $R$ to the form of modified $f(R)$ gravity
\beq
S = \frac{M^2_P}{2}\int d^4x \sqrt{-g} f(R).
\label{fRgen}
\eeq
We shall describe it here in more details than in \cite{AJV}. After that the analysis of consequences for inflation, for the odd terms in the action of original scalar, becomes trivial and can be done either directly in the physical (Jordan) frame or, after the conformal transformation, in the Einstein frame, see e.g. \cite{Oliver2018} where this procedure is used for a wide class of models.

Let us start with the potential (\ref{Vcla}) and, as in the previous section,  assume that the main non-minimal term
$\frac{\xi}{2}R\ph^2$ and the interacting
term $\frac{\la}{4!}\ph^4$ are dominating over other terms,
which are regarded small corrections. The effects
of these small terms using perturbations. The
kinetic term in the classical action of scalar field $\ph$, will be
simply neglected. This approximation corresponds to the part of
the inflationary epoch, when the potential term dominates. As we
shall see in what follows, this approximation provides the mapping
of the scalar potential to the $R+cR^2$ action with a sufficiently
large coefficient $c$. The known fact is that this theory fits well
with the observations, justifying the approximation.

Without the kinetic term\footnote{Taking it into account leads to
the non-localities, which were discussed in \cite{Spont,PoImpo}.},
the equation for the scalar field follow from Eq.~(\ref{Vcla}),
\beq
V'(\varphi)
\,=\,
m^2\varphi + \frac{\lambda}{6}\varphi^3
+ \frac{g}{2}\varphi^2 - \tau + {\tilde g} R - \xi\varphi R\,=\,0.
\label{Vequa}
\eeq

Let us solve Eq.~(\ref{Vequa}) perturbatively, in the first order in
the small parameters  $\tau$, ${\tilde g}$ and $g$. It is useful to
separate the potential in two parts,
$V(\varphi)=V_0(\varphi)+V_1(\varphi)$, where
\beq
V_0(\varphi)
&=&
\frac12 m^2\varphi^2 + \frac{\lambda}{4!}\varphi^4
- \frac{1}{2} \xi R\varphi^2,
\nn
\\
V_1(\varphi)
&=&
\frac{g}{3!}\varphi^3 - \tau\varphi + {\tilde g} R\varphi.
\label{V0V1}
\eeq
For the sake of  generality, we keep the mass-dependence exact
until the end of the consideration. The zero-order reduction of
(\ref{Vequa}) has the form
\beq
V'_0(\varphi)
\,=\,m^2\varphi_0 + \frac{\lambda}{6}\varphi_0^3 - \xi\varphi_0 R\,=\,0
\qquad
\Longrightarrow
\qquad
\varphi^2_0 \,=\, \frac{6}{\la} \big(\xi R -  m^2\big).
\label{V0}
\eeq
Substituting this result into the first-order equation, with
$\ph=\ph_0+\ph_1$, after a small algebra we obtain from
Eq.~(\ref{Vequa})
\beq
\ph_1
\,=\,
- \frac{3g}{2\la} + \frac{\tau  - {\tilde g}R}{2(\xi R - m^2)}.
\label{V1}
\eeq

According to the simplified version of the mapping (see, e.g.,
\cite{fREddi}), the function $f(R)$, in the first order of
perturbation theory, has the form
\beq
f(R) \,=\, R - \frac{2}{M_P^2}
\left[V_0(\ph_0) \,+\,  V_1(\ph_0) \,+\, \ph_1 V'_0(\ph_0)\right],
\label{fR-1}
\eeq
where the last term in square brackets vanishes obviously. In this way, substituting
(\ref{V0}) into the potential, we arrive at the expression
\beq
\frac{M_P^2}{2}f(R)
&=&  \frac{3}{2\la}m^4
+\left(\frac{M_P^2}{2} - \frac{3 \xi}{\la}m^2\right) R
+ \frac{3\xi^2}{2\la}R^2
\nonumber
\\
&& +\, \sqrt{ \frac{6}{\la}(\xi R - m^2) }\Big[
-\frac{g}{\la}(\xi R - m^2)
+ \tau
- \tilde g R \Big].
\label{fR-full}
\eeq
The first term in this expression is the induced cosmological
constant, the second is the Einstein-Hilbert term with an induced
correction to the gravitational constant, c.f.~Eq.~(\ref{shift}),
which is irrelevant under the condition $\xi m^2 \ll \la M_P^2$
mentioned above. The third term is the $R^2$, which is an important
element of the inflationary model of \cite{star}. According to the
standard evaluation \cite{star83}, the magnitude of the coefficient
$\frac{3\xi^2}{2\la}$ should be approximately $5\times 10^8$;
hence, the natural value of the main non-minimal parameter is $\xi \sim 10^4$. In the inflationary regime,
$\xi R \gg \frac{\la M_P^2}{\xi} \gg m^2$. Since $\lambda\ll 1$, this coefficient of the $R^2$ term in the effective $f(R)$ representation of the original model~({\ref{Vcla}) (valid during inflation only) much exceeds quantum-gravitational loop correction to it due to scalar fields strongly non-minimally coupled to gravity which is of the order of 
$\xi^2$ up to a logarithmic multiplier~\cite{Avr95,Ema19}. The imaginary part of this correction just determines the decay rate of the scalaron into pairs of such scalar particles and antiparticles after the end of inflation~\cite{star81}, that is the most effective channel of reheating in the Starobinsky model.\footnote{Note that we do not agree with the statement in \cite{Ema19} about the appearance of a new scalar degree of freedom in this system different from that existing in any scalar-tensor gravity, or in $f(R)$ gravity. In fact, the effective scalar particle in the Starobinsky model (scalaron), in our model (\ref{Vcla}) and its $f(R)$ representation during inflation (\ref{fR-full}) is the same. Its mass at the end of inflation is $M\sim M_P\sqrt{\lambda}\xi^{-1}$.}

All odd parameters and $m^2$ are small and their effect on the cosmic perturbations should be considered in the linear
approximation. As we are interested in the odd terms, we can
safely set $m^2$ to zero in Eq.~(\ref{fR-full}). In this way,
we arrive at the action
\beq
S= \int d^4x \sqrt{-g} \bigg\{
\frac{M^2_P}{2}R - \frac{3 \xi m^2}{\lambda}R
+ \frac{3\xi^2}{2\lambda}R^2
%% \nn \\ && \qquad
+ \,M_P\sqrt{\frac{6\xi R}{\lambda}}
\Big[\tau M^2_P
- \Big(\tilde g +\frac{g\xi}{\lambda}\Big)R \Big] \bigg\}.
\mbox{\qquad}
\label{fR2}
\eeq
As far as $\xi m^2 \ll \la M_P^2$, the second term in the integrand produces only a small shift in the inverse Newton constant,
\begin{align}
M^2_P
\,\, \longrightarrow \,\,M^2_P  - \frac{6\xi m^2}{\la} ,
\label{shift}
\end{align}
so it can be omitted and we arrive at
\beq
S=\frac{M^2_P}{2}\int d^4x \sqrt{-g} \left[ R
+\frac{\alpha}{M^2_P}R^2
+ \frac{2}{M_P}\sqrt{\frac{6\xi R}{\la}}
\left(M^2_P\tau - \beta R\right)\right],
\label{indaction}
\eeq
%%%%%%%%%%%%%%%%%%%%%%%%%%%%%%%%
%% \beq
%% S=\int d^4x \sqrt{-g} \,\bigg\{
%% \frac{1}{16\pi G}R - \frac{3 m^2 \xi}{\lambda}R
%% +\frac{3\xi^2}{2\lambda}R^2
%% + \sqrt{\frac{6\xi R}{\la}}\Big[
%% \tau - \Big(\tilde g + \frac{g\xi}{\la}\Big)R \Big] \bigg\}.
%% \label{fR}
%% \eeq
We note that, due to the odd terms in the potential (\ref{Vcla}),
the resulting function $f(R)$ has an unusual form with the
non-integer powers $\frac12$ and $\frac32$ of the scalar curvature.

The following observation is in order. It is clear that the term
proportional to the root of scalar curvature in the gravitational
action leads to an inconsistency, since in the presence of this
term there is no flat metric solution to the equations for the
metric. In the present case, this does not mean that the theory
which we are dealing with is inconsistent. Let us remember that
(\ref{fR2}) is not the fundamental action of gravity, but only the
intermediate form of a mapping of the scalar theory with the
potential (\ref{Vcla}), which is valid in the inflationary epoch
only, more precisely in the slow-role phase.

Indeed, the above effective action is valid only when the $R^2$ term dominates, and therefore the new contributions $R^{1/2}$ and $R^{3/2}$ can indeed be treated as perturbations. For low energies, one has $\xi \to 0$ and so the mapping from the potential $\eqref{Vcla}$ to the action $\eqref{fR2}$ cannot be performed. This allows us to avoid a possible disruption of the graceful exit or effects such as strong particle production or tachyonic instabilities, see e.g. \cite{He:2020ivk}.

If one aims to consider action $\eqref{fR2}$ as a fundamental theory,
valid for all $R$, then many conditions and requirements apply for
its viability, as discussed extensively e.g. in
Ref.~\cite{Appleby:2009uf}. For example, one must have $f'(R) > 0$
and $f''(R) > 0$ in order to guarantee that gravity is an attractive
force and in order to avoid ghosts, and these requirements put
constraints on the parameter space $(\xi,\lambda,\tau,\ga,g)$. In
the present case, these constraints do not apply because (\ref{fR2})
is not regarded as a fundamental action, but only as an intermediate
stage of the mapping of the scalar theory with the potential
(\ref{Vcla}) to the minimal scalar model. Furthermore,
according to the analysis of Ref.~\cite{Appleby:2009uf},
one must extend a $f(R)$ theory to negative values of $R$ in order to
guarantee a graceful exit from the inflationary case. As it stands, the
$f(R)$ theory of Eq. $\eqref{fR2}$ has not this extension because of
the $\sqrt{R}$ term, and thus can only be regarded as an effective
theory for large $R$.

%% We can rewrite the modified gravity action (\ref{fR2}) in the form

Further analysis will be based on the action (\ref{indaction}),
%%%%%%%%%%%%%%%%%%%%%%%%%%%%%%%%
that can be regarded as a particular case of the $f(R)$ theory
(\ref{fRgen}). This action can be mapped to the usual scalar-metric action (see, e.g., \cite{fREddi} and further references therein), in the Jordan frame:
\beq
S = \frac{M^2_{P}}{2}\int d^4x \sqrt{-g}
\big[ \phi R - V(\phi) \big],
\label{Vphi}
\eeq
where
\beq
\phi \,=\, f'(R) \,=\,
1 \,+\, \frac{2\alpha}{M^2_P}R
\,+\, \ga\,\bigg(\frac{\tau M_P }{{\sqrt R}}
\,-\, \frac{3\be}{M_P}\,{\sqrt R} \bigg)\;, \qquad V(\phi) = \phi R - f(R).
\label{phiR}
\eeq
Here the prime denotes derivation with respect to $R$ and
we used notation (\ref{albega}). As before, we assume all ``odd''
parameters to be small and perform all calculations perturbatively,
in the first order in these parameters.
This approach simplifies the general procedure of \cite{fREddi}.
Let us call $R_0$ the solution without odd terms and $\De R_1$
the first order correction to it. Writing $R$ as
\beq
R = R_0 + \Delta R_1
\eeq
where
$\big| \De R_1\sim O^{(1)}(\tau,\beta)\big|  \ll \big| R_0\big| $,
we arrive at
\begin{align}
R^{1/2} \approx R^{1/2}_0\left(1+\frac{\Delta R_1}{2 R_0}\right),
\qquad
R^{-1/2} \approx R^{-1/2}_0\left(1-\frac{\Delta R_1}{2 R_0}\right).
\end{align}
Solving Eq.~(\ref{phiR}), at the zero order we get
\beq
\phi
= 1+\frac{2\alpha}{M^2_P}R_0
\quad \Longrightarrow \quad R_0 =\frac{M^2_P}{2\alpha}(\phi-1)
\eeq
and at the first order
\beq
\Delta R_1
\,\approx \,
\frac{3 M_P\beta\gamma R^{1/2}_0 - M^3_P\tau\gamma
R^{-1/2}_0}{2\alpha}.
\eeq
The explicit form of the solution is
\begin{align}
R(\phi) = \frac{dV(\phi)}{d\phi}
= \frac{M^2_P}{2\al}(\phi - 1) + \frac{3\sqrt{2}M^2_P}{4}
\frac{\beta\ga }{\al^{3/2}}(\phi-1)^{1/2}
-  \frac{\tau\ga M^2_P}{\sqrt{2\al}}(\phi-1)^{-1/2}.
\end{align}
Finally, after integration, we obtain the potential
\beq
V(\phi)
\,=\,
\frac{M^2_P}{4\al}(\phi - 1)^2 + \frac{\sqrt{2}M^2_P}{2}
\frac{\be\ga }{\al^{3/2}}(\phi-1)^{3/2}
- \frac{ 2\tau\ga M^2_P}{\sqrt{2\al}}(\phi-1)^{1/2}.
\eeq
It is useful to work with the action of the standard form,
in the Einstein frame:
\begin{align}
S = \int d^4x \sqrt{-g} \bigg\{
\frac{M^2_P}{2} R
- g^{\mu\nu} \pa_\mu\chi  \, \pa_\nu \chi - U(\chi)\bigg\}.
\end{align}
Since the terms originating from the odd terms in Eq.~(\ref{Vcla})
contribute only to the potential part of the action, after the Weyl
transformation we get the usual relation between the field $\chi$
with canonically normalized kinetic term and the scalar $\phi$,
\begin{align}
U(\chi) = \frac{M^2_P}{2\phi^2}\,V(\phi(\chi)),
\qquad
\mbox{where}
\qquad
\phi(\chi) = e^{\sqrt{\frac{2}{3}}\frac{\chi}{M_P}}.
\end{align}
After a small algebra, we find for the potential the expression
(\ref{potentialUchi}). In the $m^2 = 0$ approximation, it boils
down to
%%   DO SIH
\beq
U(\chi)
&=&
\frac{M^4_P}{8\alpha}\sigma^2B^2
+ \frac{\sqrt{2}M^4_P}{4}\frac{\beta\gamma}{\alpha^{3/2}}
\,\si^2 \,B^{3/2}
- \frac{M^4_P\tau\gamma }{\sqrt{2\alpha}}\,\si^2\,B^{1/2},
\label{potential}
\eeq

%%
%% \vskip 20mm
%% \beq
%% U(\chi)
%% &=&
%% \frac{M^4_P\lambda}{24\xi^2}\sigma^2 B^2
%% + \frac{M^4_P}{6}\left(\tilde g +\frac{g\xi}{\lambda} \right)
%% \frac{\lambda }{\xi^{5/2}}\, \si^2 B^{3/2}
%% \nn \\ &-&
%% - \frac{M^4_P\tau}{\sqrt{\xi}} \, \si^2 B^{1/2}.
%% \qquad
%% \label{potentialorig}
%% \eeq

The two new terms $R^{1/2}$ and $R^{3/2}$, with the corresponding
two corrections in the above potential, therefore modify the dynamics
of inflation. Which of the two dominates depends on the parameters, but
if these are of the same order, then the $R^{3/2}$ term dominates.
Gravitational vacuum polarization from massive fermions (i.e. with
masses $m \gg H$) during inflation provides a contribution
$\sim R^3/M_P^2$ \cite{AJV} (see also \cite{AJV-2} and also
\cite{book} for more examples), but, owing to the Planck suppression,
this is small compared to the new terms due to fermions. See also
Ref.~\cite{Netto:2015cba} in connection with RG corrections to
$\xi$ resulting in its running which transforms to the running of the
$R^2$ coefficient in the $f(R)$ representation.

%%%%%%%%%%%%%%%%%%%%%%%%%%%%%%%
%%%%%%%%%%%%%%%%%%%%%%%%%%%%%%%
%%%%%%%%%%%%%%%%%%%%%%%%%%%%%%%
\section{Derivation of the slow-roll parameters}
%% in the $R+R^2$ model}
\label{s3}

In the previous sections we saw that the potential (\ref{potential})
in the Einstein frame can be derived from the original potential
(\ref{Vcla}) in two different (albeit ultimately equivalent) ways.
Now we are in a position to use this expression to obtain the
parameters characterizing the inflation.

It is useful to derive the first and second derivatives of the
potential, which have the form
\beq
U'(\chi)
&=&
\frac{\beta \gamma M^3_P \si B^{1/2}}{4\sqrt{3}\al^{3/2}}
\left(3 - 4\si B \right)
+ \frac{\sqrt{2}M^3_P\,\si^2 B}{4\sqrt{3}\al}
- \frac{\tau\ga M^3_P}{2\sqrt{3\al}  B^{1/2}}\left(\si - 4\si^2 B \right),
%%%%%%%%%%%%%%%%%%%%%%%%%%%%%%%%%
\nn
\\
U''(\chi)
&=&
\frac{M_P^2}{6\alpha} \left(2\si^2 - \si \right)
+ \frac{\sqrt{2}\beta \gamma M^2_P}{6\sqrt{\al^3 B}}
\Big( 4\si^2 B^2 - \frac{9}{2}\si B + \frac{3}{4 } \Big)
\nn
\\
&-&
\frac{\sqrt{2}M^2_P\tau\ga}{3\sqrt{\al B^3}}
\Big(4 \si^2 B^2 -\frac{3\si B}{2} - \frac{1}{4 } \Big).
\eeq
As in the general case (see e.g. \cite{Oliver2018}),
in the slow-roll approximation the slow-roll parameters are
related to the scalar field potential as follows (see e.g.
\cite{Piattella:2018hvi}):
\beq
\epsilon = \frac{M^2_P}{2}
\bigg[\frac{U'(\chi)}{U(\chi)}\bigg]^2,
\qquad
\eta =  \frac{M^2_P\,U''(\chi)}{U(\chi)}.
\label{ep_and_na}
\eeq
Keeping only the $O^{(1)}(\tau,\beta)$ terms, we get
\beq
\epsilon &=&
\frac{4}{3}\,B^{-2}
+ \frac{4\sqrt{2}}{3}\frac{\beta \ga \left(3 - 4 \si B \right)}
{\si \sqrt{\alpha B^5}}
- \frac{8 \tau\ga \sqrt{2\al}\left(1  - 4\si B \right)}{3\si B^{7/2}} ,
\label{epeta}
\\
\eta &=&
\frac{4(2\si - 1)}{3B^2\si}
+ \frac{\sqrt{2}\beta \ga \left(16 B^2\si^2 -18B \si + 3 \right)}
{3\si^2 \sqrt{\al B^5}}
%% \nonumber \\ &-&
- \frac{2 \tau\ga \sqrt{2\al}\left( 16B^2\si^2 - 6\si B - 1\right)}
{3\si^2 \sqrt{B^7}}.
\nn
\eeq
Deep in the inflationary regime, that is, for large values of $\chi$,
one can take only the leading term in each expression,
\beq
&&
\epsilon
\,\simeq\,
\frac{4}{3} e^{-2\sqrt{\frac{2}{3}}\frac{\chi}{M_P}}
- \frac{4\sqrt{2}}{3}\frac{\beta \ga}{\sqrt{\al}}
e^{-\frac{3}{2}\sqrt{\frac{2}{3}}\frac{\chi}{M_P}}
+ 8\sqrt{2}\tau \gamma \sqrt{\alpha}e^{-\frac{5}{2}
\sqrt{\frac{2}{3}}\frac{\chi}{M_P}},
\nn
\\
&&
\eta
\, \simeq \,
-\frac{4}{3} e^{-\sqrt{\frac{2}{3}}\frac{\chi}{M_P}}
+ \frac{\sqrt{2}}{3}\frac{\beta \gamma}{\sqrt{\alpha}}
e^{-\frac{1}{2}\sqrt{\frac{2}{3}}
\frac{\chi}{M_P}} - 6\sqrt{2}\tau \gamma \sqrt{\alpha}e^{-\frac{3}{2}\sqrt{\frac{2}{3}}\frac{\chi}{M_P}},
\eeq
or
\beq
\epsilon
&=&
\frac{4}{3}\sigma^2
- \frac{4\be\ga \sqrt{2\si^3}}{3\sqrt{\alpha}}
+ 8\tau \ga \sqrt{2\al \sigma^5}.
\label{epsilon}
\\
\eta
&=&
- \frac{4\si}{3}
+ \frac{\be \ga\sqrt{2\si}}{3\sqrt{\al}}
- 6\tau \ga \sqrt{2\al \si^3} .
\label{eta}
\eeq
To calculate the number of e-folds, we express $\ep$ as a sum
$\ep = \ep_0 + \delta\ep_1$,  where $\ep_0 \sim O^{(0)}(\tau,\beta)$
and $\delta\epsilon_1 \sim O^{(1)}(\tau,\beta)$,
\beq
N(\chi) \,=\,
\frac{1}{M_P}\int_{\chi_{end}}^{\chi}
\frac{d\chi'}{\sqrt{2\epsilon(\chi')}}.
\eeq
Expanding the square root in the integrand, leads to
\beq
N(\chi)
&=&
- \frac{3}{4}\int_{\sigma_{end}}^{\sigma}
\frac{d\sigma'}{{\sigma'}^{2}}
\bigg( 1
+  \frac{\be\ga}{\sqrt{2\al {\si'}}}
- 3\tau \ga \sqrt{\frac{2\al}{{\sigma'}}}
\bigg),
\label{Nchi}
\eeq
where
\beq
d\sigma' \,=\,
-\,\sqrt{\frac{2}{3}}\,\frac{\sigma' d\chi'}{M_P}.
\eeq
After integration and using the condition
$\chi_{end}\ll \chi$ ({\it i.e.} assuming that $\chi$ is deep
in the inflationary era), we find
\beq
N(\chi) =\frac{3}{4}
\left[ \sigma^{-1}
+ \frac{\sqrt{2}\be \ga}{3\sqrt{\al \si^3}}
- 6\sqrt{2\alpha}\tau \gamma \sigma^{-1/2} \right]
\label{N}.
\eeq
We can expand on $\sigma$ in Eq.~\eqref{N} to find the field $\chi$
in terms of the number of e-folds:
\begin{align}
\sigma = \sigma_0 + \delta\sigma_1
= \frac{3}{4N} + \frac{\sqrt{2}}{3}
\frac{\beta \gamma}{\sqrt{\alpha}}
\left(\frac{3}{4N}\right)^{1/2} - 6\sqrt{2\alpha}\tau \gamma
\left(\frac{3}{4N}\right)^{3/2}
\end{align}
Then, by plugging it in Eqs.~\eqref{epsilon},\eqref{eta}, and
keeping only the terms up to $O^{(1)}(\tau,\beta)$, we get
\begin{align}
\epsilon = \frac{3}{4N^2} - \frac{\sqrt{6}}{6}
\frac{\beta \gamma}{\sqrt{\alpha}}
\frac{1}{N^{3/2}} - \frac{9\sqrt{6}}{4}\tau \gamma
\sqrt{\alpha}\frac{1}{N^{5/2}},
\end{align}
\begin{align}
\eta = -\frac{1}{N}
- \frac{\sqrt{6}}{18}\frac{\beta \gamma}{\sqrt{\alpha}}
\frac{1}{N^{1/2}} + \frac{3\sqrt{6}}{4}\tau \gamma
\sqrt{\alpha}\frac{1}{N^{3/2}}.
\end{align}
The main inflationary observables are the scalar spectral index
$n_s$ and the tensor-to-scalar ratio $r$ (see e.g.
\cite{Piattella:2018hvi}), whose expressions in terms of the
slow-roll parameters are:
\begin{align}
n_s -1 = -6\epsilon + 2\eta,\qquad r=16\epsilon.
\end{align}
These observables have constrained by the Planck mission
\cite{Akrami:2018odb} to $n_s = 0.9649 \pm 0.0042$ at 68\% CL
and $r_{0.002} < 0.056$ at 95\% CL.

Therefore, keeping the leading order of the slow-roll approximation only, we arrive at
\beq
&&
n_s - 1
= - \frac{2}{N}
%\Big( 1 + \frac{9}{4N} \Big)
- \frac{\sqrt{6}\beta \gamma}{9\sqrt{\al N}}
%\Big( -1 + \frac{9}{N} \Big)
+ \frac{3\sqrt{3 \al}}{\sqrt{2 N^3}}\tau\ga,
%\Big( 1 + \frac{9}{N} \Big),
\nn
\\
&&
r = \frac{12}{N^2}
- \frac{8\sqrt{6}\,\beta \gamma}{\sqrt{3 \alpha}N^{3/2}}
- \frac{36\sqrt{6\al} \tau \ga}{N^{5/2}},
\eeq
and in terms of the original parameters:
\beq
&&
n_s - 1 = -\frac{2}{N}
%\Big( 1 + \frac{9}{4N} \Big)
- \frac{2\sqrt{3}}{9\,\sqrt{\xi N}}\Big(\tilde g +\frac{g\xi}{\lambda} \Big)
%\Big(\frac{9}{N}  -1 \Big)
+ \frac{9\sqrt{3}\,\tau \xi^{3/2}}{N^{3/2}\,\la},
%\Big( 1 + \frac{9}{N} \Big),
\label{ns}
\nn
\\
&&
r = \frac{12}{N^2} - \frac{16}{\sqrt{3 \xi N^3}}
\Big(\tilde g +\frac{g\xi}{\lambda} \Big)
- \frac{216\sqrt{3}}{N^{5/2}}\,\frac{\tau \xi^{3/2}}{\lambda }.
\label{rs}
\eeq

We have derived the above results in the Einstein frame,
but as long as we retain up to first order corrections, they are
valid also in the Jordan one, although the number of e-folds $N$ becomes a different function of the present wave vector modulus $k$. Moreover, the results (\ref{ns}) can be derived directly in the Jordan frame, without the conformal transformation to the Einstein frame, by using the formulas presented in Ref.~\cite{Liu:2018hno} where a generic $f(R) = A(R)R^2$ model, with $A(R)$ slowly varying, is explored. Note that for the case investigated in the present paper, cf. Eq.~\eqref{indaction}, the corresponding $A(R)$ function is:
\begin{equation}
	A(R) = \frac{M_P^2}{R} + \alpha
+ 2M_P\sqrt{\frac{6\xi}{\la R^3}}
\left(M^2_P\tau - \beta R\right).
\end{equation}
So, it is slowly varying for large values of $R$ for which the slow-roll inflationary regime is realized in this model.

In order to estimate the value of the inflationary observables coming
from the new terms we need the value of the parameters
$\xi,\,\la,\, \tilde g,\, g, \, \tau$. If $\lambda = 1$, we can set
$\xi \approx 1\times 10^{4}$, so that we recover Starobinsky
inflation when $\tilde g = g= \tau = 0$. The value of $\tilde g$,
$g$ and $\tau$ depends on the mass of the fermions to which the
scalar is coupled, via the renormalization group equations, as
explained in \cite{AJV}. Without repeating the corresponding
arguments, we just mention that the quantum contributions to
the dimensional parameters $\tilde g$, $g$ and $\tau$ from a
fermion loop are proportional to the respective power of the
mass of a fermion.

For the first evaluation let's assume the fermion masses at the upped
bound of the Standard Model, with $m_f \sim 1\,$TeV. Then,
according to the previous considerations, $g = \tilde g \approx 1\,$TeV and
$\tau \approx 10^{3}\,$TeV. For these values of the parameters and
$N=60$ we find:
\begin{align}
n_s = 0.965417 ,\qquad r = 0.003333,
\end{align}
and for $N = 50$:
\begin{align}
    n_s = 0.9582 ,\qquad r = 0.0048,
\end{align}
which are the same, to this precision, to the $R^2$ case.
%% \textcolor{Blue}{\LARGE \textbullet}
%%%%%%%%%%%%%%%%%%%%%%%%%%%%
%%%%%%%%%%%%%%%%%%%%%%%%%%%%
If we increase the magnitude of $m_f$ to the GUT's scale, for
$m_f \approx 10^{14}\,$ GeV we find $n_s = 0.965374$ and
$r = 0.003313$. For the supersymmetric GUT models, with
$m_f \approx 10^{16}\,$GeV, there is
$n_s = 0.961231$ (a difference of $0.43\%$ from $R+R^2$
model) and $r = 0.001332$ (a difference of $60\%$ from
$R+R^2$ model).

We can compare the contribution to the inflationary observables coming from the induced action (\ref{indaction}) with the second order corrections coming from pure Starobinsky inflation. The next-to-leading order contributions to $\epsilon$ and $\eta$ are, when $N = 60$:
\begin{align}
\epsilon_{NL} &= \frac{9}{8 N^3} = 5.21\times 10^{-6},
\\
\eta_{NL} &=  -\frac{3}{2N^2} = -4.17\times 10^{-4}.
\end{align}
For $m_f \approx 1\,$ TeV we have that the corrections for
$\epsilon$ and $\eta$ from the $\sqrt{R}$ and $R^{3/2}$ terms, when
$N=60$, are $\epsilon_{\text{odd}} = -1.24\times 10^{-17}$ and
$\eta_{\text{odd}} = -2.48\times 10^{-16}$, and are indeed smaller than the
next-to-leading order corrections of the $R+R^2$ model. However, if $m_f$ is
as large as $m_f \approx 10^{16}\,$GeV, the new contributions to $\epsilon$ and
$\eta$ are $\epsilon_{\text{odd}} = -1.25 \times 10^{-4}$ and
$\eta_{\text{odd}} = -2.47 \times 10^{-3}$ become more relevant than the
second order corrections for the Starobinsky model.

In figure~\eqref{fig:potential} we can see the potential of
Eq.~\eqref{potential} for a few different magnitudes of $m_f$,
together with the pure Starobinsky model.  For negative values of
the field, our potential becomes imaginary because of the square
root. But inflation corresponds to $R>0$ and the plateau, so in
principle this does not represent a problem.
\\
%%%%%%%%%%%%%%%%%%%%    Figure
\begin{figure}[ht]
\centering
\includegraphics[scale=0.62]{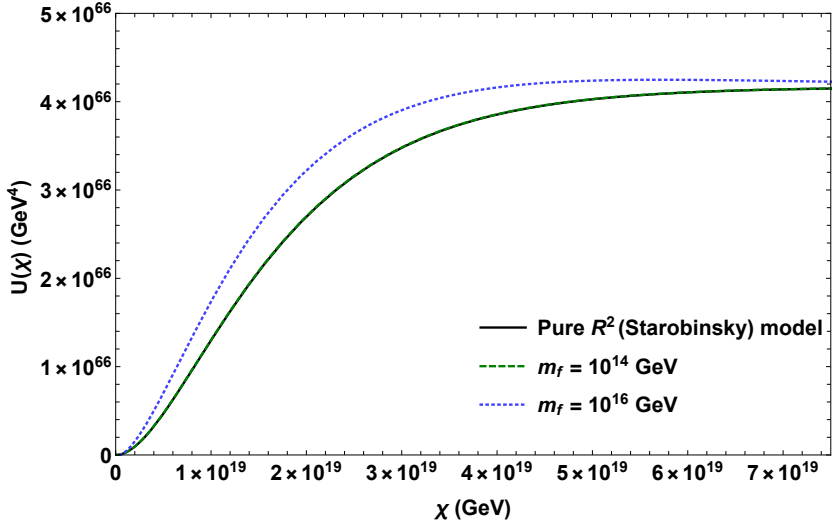}
\begin{quotation}
\caption{\sl Potentials for large values of typical masses $m_f$,
together with the reference plot for the $R^2$-model
without odd terms. In all cases, $\xi=10^{4}$. }
\end{quotation}
\label{fig:potential}
\end{figure}

\begin{figure}[ht]
\centering
\includegraphics[scale=0.62]{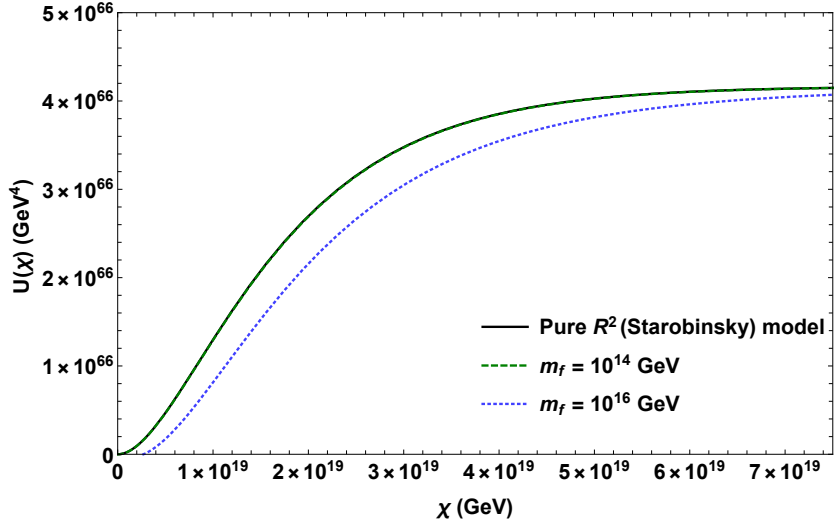}
\begin{quotation}
\caption{\sl Potentials for different large values of typical masses
$m_f$, when $g<0$ and $\xi=10^4$, and the reference plot for the
$R^2$-model without
odd terms. Note that the odd-terms effect is almost invisible for
$m_f = 10^{14}\,GeV$.}
\end{quotation}
\label{fig:potentialgneg}
\end{figure}

\begin{figure}[ht]
\centering
\includegraphics[scale=0.62]{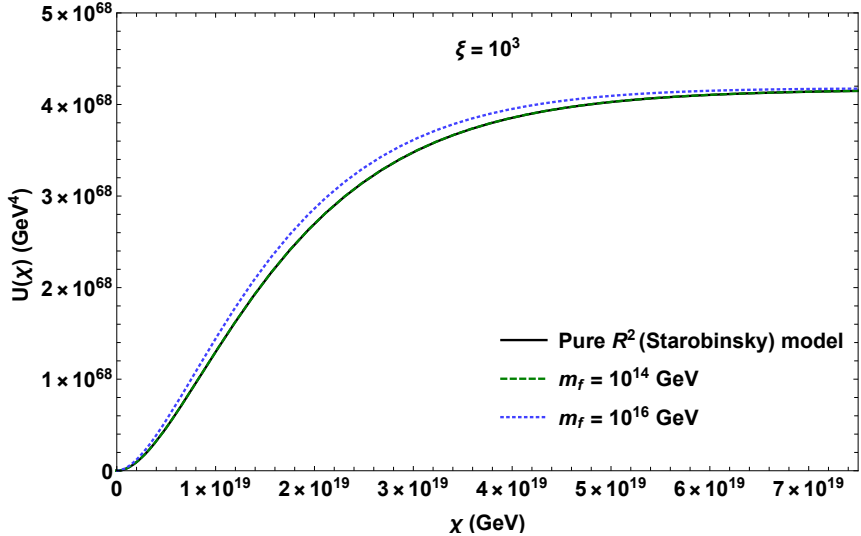}
\begin{quotation}
\caption{\sl Potentials for different large values of the mass
$m_f$, together with the reference plot for the Starobinsky
model without odd terms, $\xi=10^{3}$. }
\end{quotation}
\label{fig:potentialxi10-3}
\end{figure}

\begin{figure}[ht]
\centering
\includegraphics[scale=0.62]{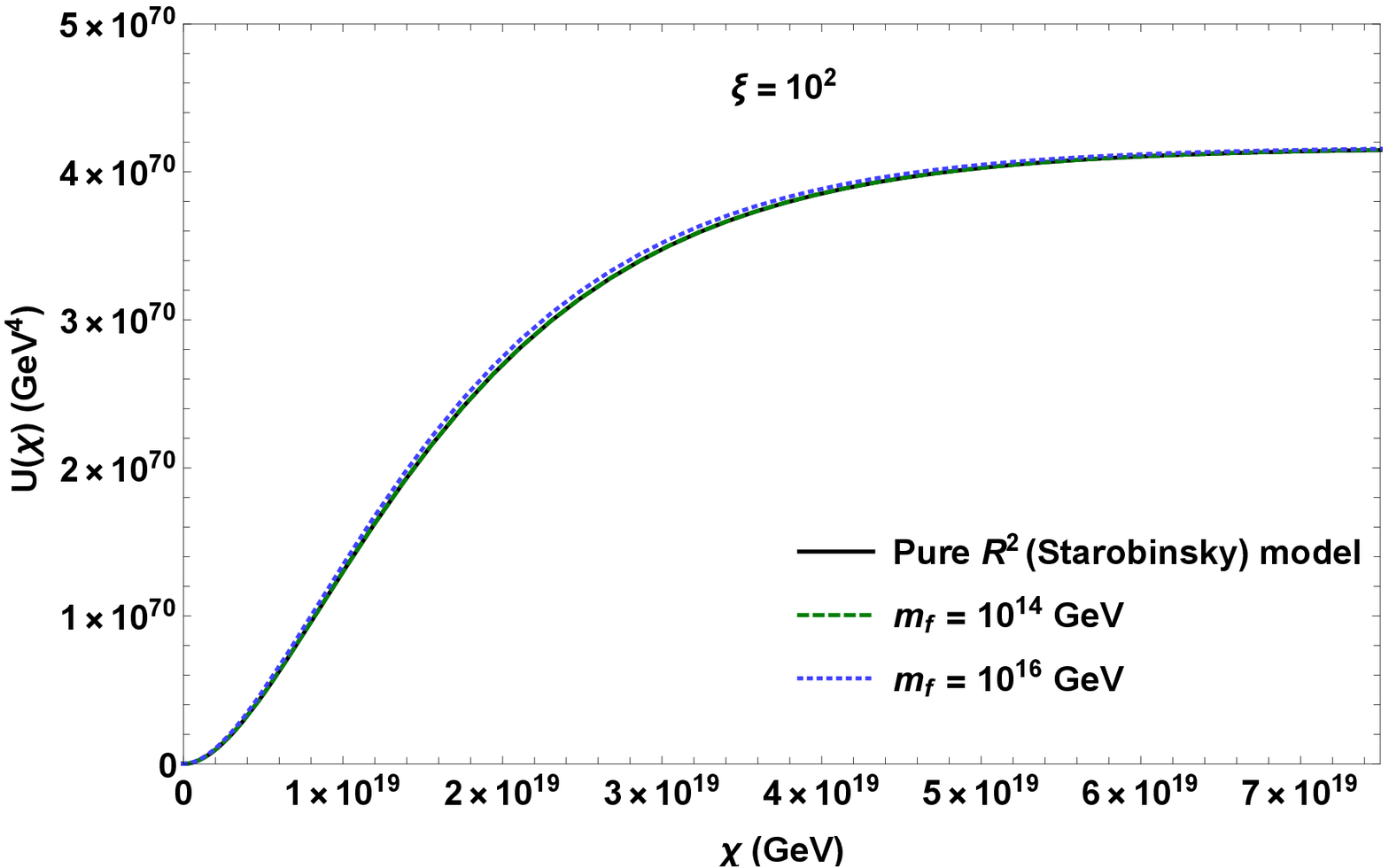}
\begin{quotation}
\caption{\sl Potentials for  large values of typical masse $m_f$,
together with the reference plot for the Starobinsky model without
odd terms, where $\xi=10^{2}$. }
\end{quotation}
\label{fig:potentialxi10-2}
\end{figure}

In order to understand better the results, at this point we have to
come back to the definition of the problem in \cite{AJV} and
describe the physical situation in which the sterile scalar can be
coupled to the Standard Model fermions.

The Standard Model left-handed fermions are doublets under the
$SU(2)$ gauge group of the form
$\begin{pmatrix} u^L_i \\ d^L_i \end{pmatrix}$ in the case of
left-handed quarks. We have $u^L_i=\{u_L, t_L, c_L \}$ and
$d^L_i = \{d_L,b_L,s_L\}$. On the other hand, the right-handed
fermions are singlets under $SU(2)$ group, namely $u^R_i$ and
$d^R_i$. So, in
order to have Yukawa interaction in the Lagrangian being invariant
under the $SU(2)$ gauge group, the scalar must be at least a
doublet under $SU(2)$, so it can multiply the $SU(2)$ index of
the left-handed fermions. This is exactly why we cannot introduce
directly the fermions mass terms into the Lagrangian, and we have
to do so using the Higgs mechanism.

%% That is, we would have a term of the form (with the $SU(2)$ representations written explicitly), $\begin{pmatrix}\bar{u}^L_i \\ \bar{d}^L_i \end{pmatrix}\phi u^R_i $ in the Lagrangian, and we can see that $\phi$ should be something like $\phi = \begin{pmatrix} \phi_1 & \phi_2 \end{pmatrix}$ under $SU(2)$. The fact that we have to couple the left and right-handed fermions (instead of just something like $\bar{u}^L_i u^L_i$) is because we also need invariance under the $U(1)_{Y}$ gauge group, so we can only write terms where the hypercharge $Y$ adds to zero (that's exactly why we cannot introduce directly the fermions mass terms into the Lagrangian, and we have to do so using the Higgs mechanism).

The only possibility to couple a sterile scalar to Standard Model
fermions is when the $SU(2) \times U(1)$ symmetry is broken by
the Higgs mechanism, at energies below  125 GeV.
At this regime, terms in the effective low-energy Lagrangian don't
need to be invariant under SU(2), as this is no longer a manifest
symmetry of the theory. In particular, if the sterile scalar is mixed
with the Higgs at high energies, in the process of symmetry breaking
the Yukawa interactions with the sterile scalar emerge in a natural way.

%% However, doing so always will lead to the problem of what happens to this scalar field at high energies, as it cannot couple to the fermions anymore. I think that in ``standard inflationary models'' what happens is that the reheating (decaying of inflaton into SM particles) is done through the SM bosons first, and the bosons decay to the fermions afterwards, so we don't have an interaction between them and the inflaton (as far as I know, but I can be wrong!) } Therefore, we can think of physics beyond the Standard Model, where we could have new heavy singlets fermion fields that could couple to a sterile scalar. So right-handed neutrinos could  be a possibility, as well as GUT's (I guess).

On the other hand, as we have seen above, the coupling of a sterile
scalar (inflaton) with fermions at the Standard Model energy and mass
scale does not produce essential changes in the inflationary observables.
If thinking about the physics beyond the Standard Model, there may be
new heavy fermion singlet fields, that could couple to a sterile scalar.
Alternatively, the coupling of the inflaton with the Higgs-like scalar
of GUT model can give the effect of mixing similar to the one
described above for the Standard Model. This possibility gives a
chance to detect the traces of GUT's in the cosmological observations.

%%%%%%%%%%%%%%%%%%%%%%%%%%%%%%%%%%
%%%%%%%%%%%%%%%%%%%%%%%%%%%%%%%%%%
%%%%%%%%%%%%%%%%%%%%%%%%%%%%%%%%%%
\section{Conclusions}
\label{sec5}

We have explored basic consequences of odd terms in the inflaton potential in the case when the inflaton is strongly non-minimally coupled to gravity. The presence of these odd
terms is motivated  by the structure of renormalization of a
generic sterile scalar coupled to fermions by means of the Yukawa interaction. The analysis has been performed both by a direct transformation to the Einstein frame, and also by means of mapping to the $f(R)$ inflationary model in the original Jordan frame which has the same predictions for primordial perturbation spectra. Along with an additional control, this second approach provides more intuitive understanding of the role of the odd terms.

An advantage of our approach is that the physical analysis can be
performed in terms of the underlying particle physics model to which
the inflaton is coupled. The values of the constants of the odd terms
in the potential satisfy the lower bounds related to the running of
the corresponding parameters. In practice, this means these
dimensional constants should be of at least the same order of
magnitude as the heaviest fermions of the model.  The main result
which we obtained is that the effect of the odd terms is negligible
of the typical mass $m_f$ of the heaviest fermions is smaller than
the GUT scale about $10^{16}\,$GeV. Thus, the odd terms in the
inflaton potential become relevant only in the presence of a GUT
with the corresponding fermions. If these conditions are satisfied,
there is, in principle, a chance to distinguish the inflaton models
from the legitimate $f(R)$ models by measuring the quantities
such as the spectral index $n_s$ and tensor-to-scalar ratio $r$.

%%%%%%%%%%%%%%%%%%%%%%%%%%%%%%%%%%
%%%%%%%%%%%%%%%%%%%%%%%%%%%%%%%%%%
%%%%%%%%%%%%%%%%%%%%%%%%%%%%%%%%%%
\section*{Acknowledgments}
The authors are grateful to A.~Belyaev and R.~D.~Matheus for the
discussion of the possibility to couple sterile scalar to the Standard
Model fermions. OFP thanks CNPq, CAPES and FAPES for
partial financial support.
The work of ISh was partially supported by Conselho Nacional de
Desenvolvimento Cient\'{i}fico e Tecnol\'{o}gico - CNPq under the
grant 303635/2018-5 and by Russian Ministry of Science and High
Education under the project No. FEWF-2020-0003.
AAS was partly supported by the project number 0033-2019-0005 of the Russian Ministry of Science and Higher Education.

%%%%%%%%%%%%%%%%%%%%%%%%%%%%%%%%%%%%%%%

%%%%%%%%%%%%%%%%%%%%%%%%%%%%%%
\bibliographystyle{unsrturl}
%\bibliography{References}

\end{document}